# A New Clustering Algorithm Based on Near Neighbor Influence

Xinquan Chen[1, 2]

[1]Web Sciences Center, University of Electronic Science & Technoloughy of China, China

[2]School of Computer Science & Engineering, Chongqing Three Gorges University, China

chenxqscut@126.com

**Abstract**: This paper presents Clustering based on Near Neighbor Influence (CNNI), a new clustering algorithm which is inspired by the idea of near neighbor and the superposition principle of influence. In order to clearly describe this algorithm, it introduces some important concepts, such as near neighbor point set, near neighbor influence, and similarity measure. By simulated experiments of some artificial data sets and seven real data sets, we observe that this algorithm can often get good clustering quality when making proper value of some parameters. At last, it gives some research expectations to popularize this algorithm.

**Keywords**: dissimilarity measure; near neighbor point set; near neighbor influence; clustering

## 1. Introduction

Clustering is an unsupervised learning method which tries to find some distributions and patterns in unlabeled data sets. Usually, those points in the same cluster should have more similarity than other points in other clusters. There have been thousands of clustering algorithms until now, but none of them is all purpose.

Clustering algorithms are usually categorized into partitioning methods, hierarchical methods, density-based methods, grid-based methods, and model-based methods [8, 13].

KMeans [11] is a classical partitioning clustering algorithm presented in 1967. It has some inherent shortcomings. For example, this kind of algorithm needs a predefined number of clusters, but this number cannot be obtained correctly before clustering process. KMeans only adapts to find those globular clusters with near radius because of its assignment strategy of point to the nearest cluster. When the number of clusters is large, KMeans is very sensitive to initialization.

Affinity Propagation (AP) [5] is a new type of clustering algorithm published on *Science* in 2007. It needs a predefined similarity measure and $\Theta(n^2)$ time and space cost when computing and storing a similarity matrix corresponding to its data set. Finally, it produces a clustering result but can't give a hierarchical clustering structure in a form of multi-level tree.

This paper presents a new Clustering algorithm based on Near Neighbor Influence (CNNI), which is inspired by the idea of near neighbor and the superposition principle of influence.

The remainder of this paper is organized as follows. Section 2 lists several related papers. Section 3 gives some basic concepts. Section 4 presents a new clustering algorithm based on near neighbor influence. Section 5 validates the algorithm by some simulated experiments. Conclusions and future works are presented in Section 6.

## 2. Related Work

This paper is related in some aspects to several papers [2, 6, 7, 9, 12], although it was developed quite independently. Jarvis and Patrick [9] presented a shared nearest neighbor approach to similarity first. A similar idea was later proposed in ROCK [6]. In [9], a shared nearest neighbor graph is constructed from a proximity matrix, in which a link is created between a pair of points $X$ and $Y$ if and only if point $X$ and point $Y$ have each other in their closest $k$ nearest neighbor lists.

L. Ertöz et al. [2] presented an improved J-R clustering method by redefining the similarity between pairs of points in terms of how many nearest neighbors the two points share.

This paper is also superficially similar to some density-based clustering methods, such as DBSCAN [3]. DBSCAN needs two parameters, and our CNNI algorithm only needs one parameter that can be determined easily. You will see that they have essential difference in basic idea and concrete implementations.

## 3. Some Basic Concepts

Suppose there is a data set $S = \{X_1, X_2, \ldots, X_n\}$ in an $m$-dimensional Euclidean space. Naturally, we can use Euclidean metric as our dissimilarity measure. In order to describe our algorithm clearly, some concepts are presented first.

**Definition 1**. The $\delta$ near neighbor point set $\delta(P)$ of point $P$ is defined as

$$\delta(P) = \{X \mid 0 < d(X, P) \leq \delta, X \in S\}, \tag{1}$$

where $d(X,P)$ is the dissimilarity measure between point $X$ and point $P$ in the data set $S$. $\delta$ is a predefined threshold parameter.

**Definition 2.** One kind of similarity measure based on dissimilarity measure is defined as

$$s(X, P) = 1/(1+d(X, P)), \qquad (2)$$

where $s(X, P)$ is the similarity measure between point $X$ and point $P$, and $d(X, P)$ is the dissimilarity measure between point $X$ and point $P$. When $X = P$, there are $d(X, P) = 0$ and $s(P, X) = 1$. So we can ensure $0 < s(X, P) \leq 1$.

In this paper, we use equation (2) to translate the dissimilarity measure to the similarity measure. In some cases, $s(X, P) = \exp(-d(X, P))$ can also be used as a kind of similarity measure [12]. For some local clustering problems, perhaps $s(X, P) = \exp(-d(X, P) / \delta)$ is a better similarity measure.

**Definition 3**. The $\delta$ near neighbor influence $I_\delta(P)$ of point $P$ is defined as

$$I_\delta(P) = \sum_{X \in \delta(P)} s(X, P), \qquad (3)$$

where $\delta(P)$ is the $\delta$ near neighbor point set of point $P$, and $s(X, P)$ is the similarity measure between point $X$ and point $P$.

The concept of near neighbor influence is presented based on the superposition principle of influence. It is similar to the influence function [2] and the notion of density [7]. In [2], the sum of the similarities of a point's nearest neighbors is taken as a measure of this density.

## 4. Clustering based on Near Neighbor Influence

In order to implement this algorithm, first we should construct the $\delta$ near neighbor point sets of the data set $S = \{X_1, X_2, \ldots, X_n\}$ basing on one kind of dissimilarity measure. Then sort $S$ with descending sequence basing on the $\delta$ near neighbor influence of every point. Finally, basing on an idea of boosting-spread clustering, the sorted data set can be analyzed step by step by disjoint-set structure. This new clustering method is termed CNNI algorithm.

### 4.1 The Description of CNNI Algorithm

**Algorithm Name**: Clustering based on Near Neighbor Influence (CNNI)

**Input**: data set $S = \{X_1, X_2, \ldots, X_n\}$, dissimilarity measure $d(\cdot, \cdot)$, parameters $\delta$.

**Output**: a cluster label array Label[1..$n$] of the data set $S$.

**Procedure**:

Step1. First construct the $\delta$ near neighbor point set $\delta(X_i)$ for each point $X_i$ ($i = 1, 2, \ldots, n$) according to Definition 1. Then compute the $\delta$ near neighbor influence $I_\delta(X_i)$ for each point $X_i$ ($i = 1, 2, \ldots, n$) according to Definition 3.

Step2. Sort the data set $S = \{X_1, X_2, \ldots, X_n\}$ with descending sequence basing on $I_\delta(X_i)$ of every point $X_i$ ($i = 1, 2, \ldots, n$). Without loss of generality, suppose there is

$I_\delta(X_1) \geq I_\delta(X_2) \geq \ldots \geq I_\delta(X_n)$.

Step3. Construct a disjoint-set for every point in the data set $S$. The constructing procedure is listed as below:

    **for** ($i = 1; i \leq n; i++$)

        Make_Set($X_i$);   /* Make_Set(X) is an operation of the disjoint-set data structure.*/

Step4. Construct a structure DS($X_i$, $\delta(X_i)$) for every point $X_i$ ($i = 1, 2, \ldots, n$). Here, DS($X_i$, $\delta(X_i)$) is composed of a label index of point $X_i$ and its corresponding adjacency-list structure used to store $\delta(X_i)$. An array with $n$ structures can be expressed as (($X_1$, $\delta(X_1)$), ($X_2$, $\delta(X_2)$), …, ($X_n$, $\delta(X_n)$)).

Step5. Construct and merge those clusters of the data set $S$. Here, the array element Label[$i$] is used to record the cluster label of point $X_i$ ($i = 1, 2, \ldots, n$). The constructing and merging procedure is listed in Table 1.

Table 1. The concrete procedure of Step5 in CNNI Algorithm

```
1    for (i = 1; i ≤ n; i++)
2        Label[i] ← 0;   /* Every point is regarded as an isolated point at first. */
3    i ← 1;
4    j ← 1;   /* Tag j records the current number of clusters. Its initial value is set to 1. */
5    while ((j ≤ n) and (i ≤ n))
6    {
7        Cj ← Ø;   /* The j-th cluster Cj is set to Ø at first. */
8        if ((Label[i] == 0) and (many points of δ(Xi) are not assigned to any cluster Ct (0 < t < j), which is constructed before))   /* If Label[i] is equal to 0 and the Label[] value of many points in δ(Xi) is also equal to 0, then the condition of the if() sentence is true. In our experiments, "many" is set to       (0.8*|δ(Xi)|). */
9        {
10           Cj ← Union(Xi, P∈δ(Xi));   /* Union(·, ·) is an operation of the disjoint-set data structure. After this operation, the j-th cluster Cj stores point Xi and δ(Xi). */
11           Label[i] ← j;
12           Label[{l|Xl ∈ δ(Xi)}] ← j;   /* The Label[] value of all points in δ(Xi) is set to j. That means those points in δ(Xi) are assigned to the j-th cluster Cj. */
13           j ← j+1;
14       }
15       else
16           if (Label[i] == 0)
17           {
18               Point Xi joins to one cluster Cr (0 < r < j) constructed before, which contains the most points of δ(Xi);   /* Here, "most" is not the same as "many" above. It means that Cr (0 < r < j) is the cluster that contains the largest number of points of δ(Xi). In this step, we call Find_Set(X), an operation of the disjoint-set data structure, to search the cluster set of point X. */
19               Label[i] ← r;
20           }
21       i ← i+1;
22       while ((i < n) and (Label[i] ≠ 0))
23           i ← i+1;
24   }
```

Step6. Finally we get a super set $\{C_1, \ldots\}$, which is a set of clusters of the data set $S = \{X_1, X_2, \ldots, X_n\}$. The cluster label of every point in $S$ is recorded by the array

Label[1..$n$].

**4.2 Some Notes of CNNI Algorithm**

(1) Make_Set($X$), Union($X$, $Y$), and Find_Set($X$) are three basic operations of the disjoint-set data structure [10].

(2) **Time complexity analysis:**

In Step1, if we don't use any improving technique, then constructing the $\delta$ near neighbor point set of a point needs Time = $O(n)$ and Space = $O(n)$. Computing all $\delta$ near neighbor influence $I_\delta(X_i)$ ($i$ = 1, 2, …, $n$) of the data set $S$ needs Time = $O(n^2)$ and Space = $O(kn)$. Where $k = \max\{|\delta(X_1)|, |\delta(X_2)|, …, |\delta(X_n)|\}$.

In Step2, we use a sorting algorithm with $O(n\log n)$ average time cost.

Because operation Make_Set($X_i$) needs Time = $O(1)$, Step3 needs Time = $O(n)$.

Step4 needs Time = $O(kn)$ and Space = $O(kn)$.

Because operation Union($X_i$, $P \in \delta(X_i)$) needs Time = $\Omega(|\delta(X_i)|)$, Step5 needs Time = $\Omega(n + |\delta(X_1)| + |\delta(X_2)| + … + |\delta(X_n)|)$.

(3) Although we use the Euclidean metric as our dissimilarity measure in this paper, the algorithm is by no means restricted to this measure and this data space. If we can construct a proper dissimilarity measure in a hybrid-attribute space, then the algorithm can also be used.

**4.3 Setting parameter $\delta$**

Parameter $\delta$ will affect the results of clusters. In some cases, when $n$ is very large, we can make $|\delta(X_i)| \ll n$ be valid almost for all $i$ ($i$ = 1, 2, …, $n$) by setting a proper $\delta$. Intuitively, if the dissimilarity measure of two points is smaller than $\delta$, then the two points should be in the same cluster. Here, we present two methods to estimate parameter $\delta$.

(1) The first method is described by equation (4) for estimating parameter $\delta$.

$$DisInCluster_{\max} \leq \delta < DisDifferentClusters_{\min}, \qquad (4)$$

where $DisInCluster_{\max} = \max\{d(X,Y) \mid InOneCluster(X,Y), X,Y \in S\}$ is the maximal dissimilarity measure of two points that should be in the same cluster, and $DisDifferentClusters_{\min} = \min\{d(X,Y) \mid InDifferentClusters(X,Y), X,Y \in S\}$ is the minimal dissimilarity measure of two points that should be in different clusters.

(2) The second method is a concrete algorithm based on a MST of the data set $S$ (or a sample of $S$). That is:

a. First construct a MST from the data set $S$.

b. Then sort all edges in the MST. After that, we can obtain an increase edge set ES(MST($S$)) = {$e_1, e_2, …, e_{n-1}$}, where $e_1 \leq e_2 … \leq e_{n-1}$.

c. Search two near neighbor edges in ES(MST($S$)) that have the maximal diversification. Then another equation for estimating parameter $\delta$ can be represented as:

$$e_k \leq \delta < e_{k+1}, \text{ such that } k = \arg\max_{i=1,2,…,n-2}(e_{i+1} - e_i), \quad (5)$$

where $e_k$ and $e_{k+1}$ are two near neighbor edges having maximal diversification in ES(MST($S$)).

**4.4 The Improved Version of CNNI Algorithm (ICNNI)**

The main cost of CNNI algorithm is constructing the $\delta$ near neighbor point sets of the data set $S$. If we can design a proper index structure and use some improving techniques, then constructing $\delta(X_i)$ ($i$ = 1, 2, …, $n$) will cost less time with the same result. The improved version of CNNI algorithm by decreasing time complexity of constructing the $\delta$ near neighbor point sets of the data set $S$ is named as ICNNI algorithm. The detailed description and discussion on how to make some improvements in constructing the $\delta$ near neighbor point sets of the data set $S$ is given in [1].

**4.4 The Extended Version of CNNI Algorithm (ECNNI)**

The Extended version of CNNI algorithm (ECNNI) can be obtained by revising the concrete procedure of Step5 in CNNI Algorithm. Usually, parameter $\delta$ in ECNNI algorithm can be set as a smaller value than in CNNI algorithm so that less space and time are need to construct and store the $\delta$ near neighbor point set {$\delta(X_1), \delta(X_2), …, \delta(X_n)$}.

Table 2. The concrete procedure of Step5 in ECNNI Algorithm

```
1 – 7 are the same as 1 – 7 of Table 1.
8 – 14 are revised as:
8          if ((Label[i] == 0) and (many points of δ(Xi) are not assigned to any cluster Ct (0 < t < j), which is constructed before))
9          {
10             Cj ← Union(Xi, P∈δ(Xi));
11             Label[i] ← j;
12             Label[{l|Xl ∈ δ(Xi)}] ← j;
12.1           for (every point P in δ(Xi))
12.2               if (|δ(P)| > 0)
12.3                   A recursion function, PutNearNearborPointsIntoCurrentCluster (the current point P, the current cluster ID j, the data set S, the δ near neighbor point set {δ(X1), δ(X2), …, δ(Xn)}, the cluster label array Label[1..n]), is called;
13             j ← j+1;
14         }
15 – 24 are the same as 15 – 24 of Table 1.

The recursion function called in step 12.3 is defined as:
25   void PutNearNearborPointsIntoCurrentCluster(the current point P, the current cluster ID j, the data set S, the δ near neighbor point set {δ(X1), δ(X2), …, δ(Xn)}, the cluster label array Label[1..n])
26   {
27       for (every point Q in δ(P))
28           if ((Label[the index number in array Label of point Q] == 0) and (|δ(Q)| > 0))
29           {
30               Cj ← Union(Q, P);
31               Label[the index number in array Label of point Q] ← j;
32               PutNearNearborPointsIntoCurrentCluster (Q, j, S, {δ(X1), δ(X2), …, δ(Xn)}, Label[1..n]);
33           }
34   }
```

## 5. Simulated Experiments

**5.1 Experimental Design**

Our experiments are finished in a personal computer (Capability Parameters: Intel(R) Pentium(R) Dual CPU T3200 2.0GHz, 2G Memory). Experimental programs are developed using Visual C++6.0 under Windows XP.

To verify the validity and time efficiency of this algorithm, there will be some experiments of some artificial data sets, two UCI data sets and two bmp pictures in the next subsections.

Eight kinds of artificial data sets are produced in a 2-D region [0, 600] × [0, 600] by a program. Six artificial data sets are drawn by hand in a 2-D region referencing the original DBSCAN paper [3]. They are described in Table 3.

Table 3. The Description of Artificial Data Sets

(a) Eight Kinds of Artificial Data Sets produced by two functions in **Online Resource 1** of this paper.

| Data Sets (DS) | Number of Clusters (NC) | With Noise | Cluster Semidiameter (CS) |
|---|---|---|---|
| DS1 | 5 | yes | 36 |
| DS2 | 5 | no | 36 |
| DS3 | 8 or 9 | yes | 36 |
| DS4 | 8 or 9 | no | 36 |
| DS5 | 5 | no | 60 |
| DS6 | 8 | no | 40 |
| DS7 | 5 | no | 50 |
| DS8 | 11 | no | 40 |

(b) Six Artificial Data Sets drawn by hand in a 2-D region referencing the original DBSCAN paper.

| Data Sets (DS) (300 points) | Number of Clusters (NC) | With Noise |
|---|---|---|
| data1 | 2 | no |
| data2 | 3 | no |
| data3 | 4 | no |
| data4 | 4 | no |
| data5 | 4 | yes |
| data6 | 1 or 2 | no |

Iris and Wine are two UCI data sets [4] used in our experiments.

Two bmp pictures are obtained from Internet.

In subsection 5.2, using some artificial data sets, Clustering based on near neighbor influence (label: CNNI) and its improved version (label: ICNNI) will be compared with KMeans algorithm, FCM algorithm, and AP algorithm in time cost (measured by second) and clustering quality (intuitively displayed by several comparative clusters figures).

In subsection 5.3, CNNI and its extended version (label: ECNNI) will be compared with DBSCAN in clustering quality (intuitively displayed by several comparative clusters figures) by using some artificial data sets.

In subsection 5.4, CNNI will be compared with KMeans and AP in time cost (measured by second) and clustering purity by using two UCI data sets.

In subsection 5.5, the Compress Result of CNNI algorithm is validated by clustering 200 * 200 pixel points of two bmp pictures in RGB Space.

In subsection 5.6, the valid interval of parameter $\delta$ between CNNI algorithm and ECNNI algorithm are compared by using some artificial data sets.

Performance and effect of algorithms are measured by time cost (label: ST),

number of clusters (label: NC), the average deviation between all points and their means (label: ADM), comparative figures, or clustering purity [14] (label: CP).

In the following experiments, we implement the definition of data structures and three basic operations of the disjoint-set structure basing on some contents in [10].

**Note**: Some labels are listed as below.

DS: Data Sets

ST: Spending Time of clustering procedure (Time unit: second)

ADM: the Average Deviation between all points and their Means

NC: Number of Clusters of data set

CP: Clustering Purity

## 5.2 Compare with KMeans, FCM and AP Using Some Artificial Data Sets (DS1 – DS8, data1 - data6)

Since $\delta$ near neighbor point of point $P$ locates in the grid cell of point $P$ or its near grid cells, then if we set $r_i$ ($i$ = 1, 2, …, $m$) ≥ $\delta$, less time is needed to construct $n$ $\delta$ near neighbor point sets. The detailed discussion on how to construct grid cells is described in [1].

In this experiment, two parameters in CNNI and ICNNI are set to: as: $r_i$ ($i$ = 1, 2, …, $m$) = 20, and $\delta$ = 18. Where $r_i$ ($i$ = 1, 2, …,$m$) is the interval length in the $i$-th dimension of grid cell [1], and $\delta$ is the threshold parameter in Definition 1.

In Table 4, Table 5, and Figure 1 - 2, there are some comparative experimental results. Intercomparing CNNI, ICNNI, and KMeans, we find that ICNNI is faster than CNNI with the same clustering results (the same ADM, NC, and the same clusters displayed by figures). Intercomparing CNNI, ICNNI, and AP, we find that ICNNI is faster than CNNI with the same clustering results, and CNNI and ICNNI are faster than AP with similar clustering quality (displayed by Figure 1 and other figures in **Online Resource 2**).

Figure 1 and Figure 2 are two comparative figures of clustering results of a 2-D data set. All figures displayed the clusters of these data sets are presented in **Online Resource 2** of this paper.

From these comparative figures, we find that CNNI, ICNNI, and ECNNI can get better clustering quality (obvious clusters displayed by figures) than KMeans and FCM for some artificial data sets (DS1 – DS8), and ECNNI can get better clustering quality (obvious clusters displayed by figures) than KMeans and FCM for some artificial data sets (DS1 – DS8, data1 – data6). Especially, CNNI, ICNNI, and ECNNI can often easily find some noises or isolated points.

Table 4. Comparison of Clustering results of three clustering algorithms

| DS | CNNI / ICNNI | | | KMeans | | |
|---|---|---|---|---|---|---|
| | *ST* | *NC* | *ADM* | *ST* | *NC* | *ADM* |
| DS1(*n*=4000) | 3 / 1 | 15 | 13.72 | 0 | 5 | 30.03 |
| DS2(*n*=4000) | 3 / 0 | 5 | 13.76 | 1 | 5 | 29.65 |
| DS3(*n*=4000) | 3 / 1 | 26 | 15.04 | 0 | 9 | 31.56 |
| DS4(*n*=4000) | 3 / 1 | 8 | 15.11 | 0 | 9 | 31.10 |
| DS1(*n*=8000) | 12 / 3 | 14 | 13.76 | 0 | 5 | 30.32 |
| DS2(*n*=8000) | 12 / 3 | 5 | 13.77 | 0 | 5 | 188.30 |
| DS3(*n*=8000) | 11 / 3 | 31 | 15.08 | 0 | 9 | 30.81 |
| DS4(*n*=8000) | 11 / 2 | 8 | 15.13 | 0 | 9 | 30.49 |
| DS1(*n*=12000) | 27 / 7 | 12 | 13.76 | 0 | 5 | 50.87 |
| DS2(*n*=12000) | 27 / 7 | 5 | 13.77 | 1 | 5 | 50.78 |
| DS3(*n*=12000) | 26 / 5 | 21 | 15.06 | 0 | 9 | 30.81 |
| DS4(*n*=12000) | 27 / 5 | 8 | 15.07 | 0 | 9 | 30.72 |

Table 5. Comparison of Clustering results of three clustering algorithms

| DS | CNNI / ICNNI | | | AP | | |
|---|---|---|---|---|---|---|
| | *ST* | *NC* | *ADM* | *ST* | *NC* | *ADM* |
| DS1(*n*=400) | 0 / 0 | 15 | 13.30 | 117 | 13 | 11.26 |
| DS2(*n*=400) | 1 / 0 | 5 | 13.64 | 122 | 11 | 9.62 |
| DS3(*n*=400) | 0 / 0 | 30 | 12.74 | 125 | 10 | 17.28 |
| DS4(*n*=400) | 0 / 0 | 9 | 13.51 | 123 | 9 | 13.51 |
| DS1(*n*=800) | 1 / 0 | 15 | 13.70 | 1291 | 19 | 8.51 |
| DS2(*n*=800) | 0 / 0 | 5 | 13.88 | 1112 | 17 | 7.44 |
| DS3(*n*=800) | 1 / 0 | 33 | 13.75 | 1096 | 21 | 11.72 |
| DS4(*n*=800) | 0 / 0 | 9 | 14.20 | 1105 | 18 | 10.14 |
| DS1(*n*=1000) | 0 / 0 | 15 | 13.61 | 2200 | 21 | 7.76 |
| DS2(*n*=1000) | 0 / 0 | 5 | 13.76 | 2182 | 20 | 6.64 |
| DS3(*n*=1000) | 0 / 0 | 27 | 14.60 | 2151 | 24 | 9.85 |
| DS4(*n*=1000) | 1 / 0 | 8 | 14.92 | 2151 | 22 | 8.85 |
| DS1(*n*=2000) | 1 / 0 | 14 | 13.66 | 19473 | 31 | 6.17 |
| DS2(*n*=2000) | 1 / 1 | 5 | 13.72 | 19433 | 31 | 5.54 |
| DS3(*n*=2000) | 1 / 1 | 26 | 14.92 | 19323 | 36 | 7.41 |
| DS4(*n*=2000) | 1 / 1 | 8 | 15.05 | 19359 | 35 | 6.73 |

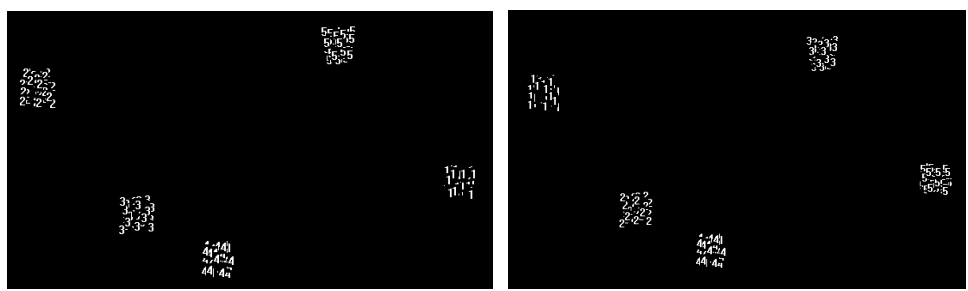

(a) Clusters identified by CNNI and ICNNI    (b) Clusters identified by AP

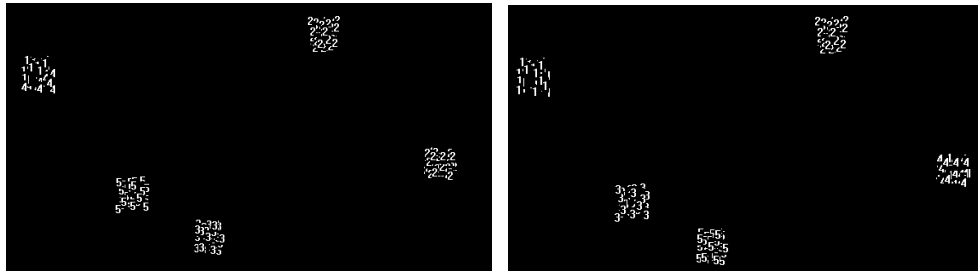

(c) Clusters identified by KMeans        (d) Clusters identified by FCM

Figure 1. Comparison of clustering results of several algorithms (DS2, *n*=200)

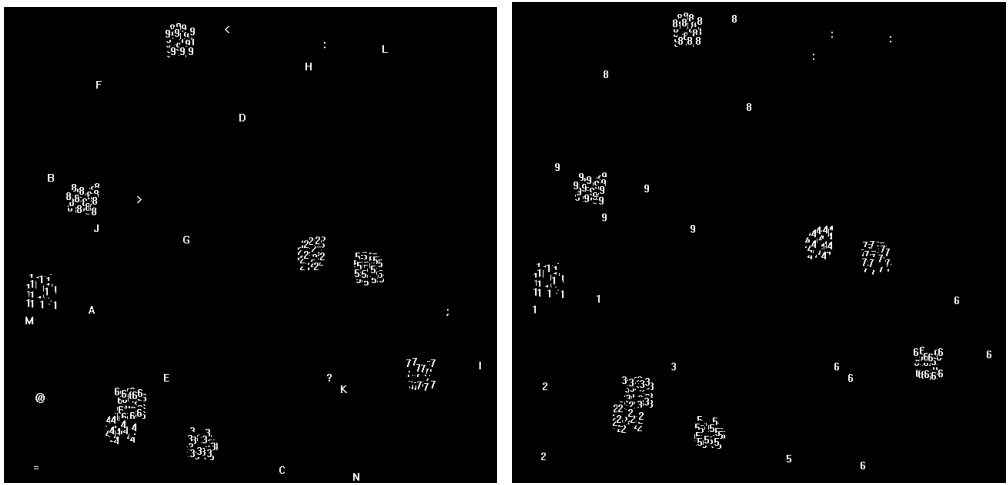

(a) Clusters identified by CNNI and ICNNI    (b) Clusters identified by AP

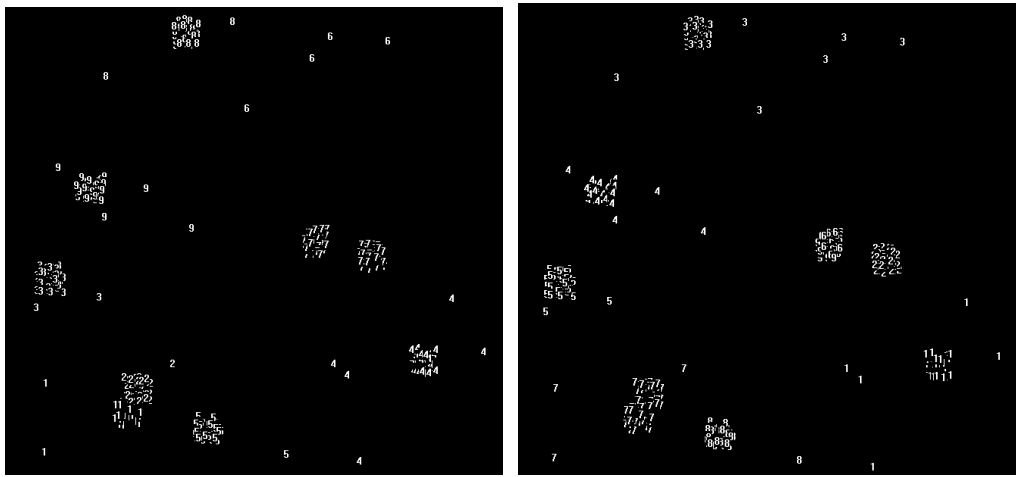

(c) Clusters identified by KMeans        (d) Clusters identified by FCM

Figure 2. Comparison of clustering results of several algorithms (DS3, *n*=400)

**5.3 Compare with DBSCAN Using Some Artificial Data Sets (data1 - data6)**

Figure 3. compares the clustering results of some artificial data sets (data1-data5) between ECNNI algorithm and DBSCAN algorithm.

Figure 4. compares the clustering results of an artificial data set (data6) among CNNI, ECNNI and DBSCAN algorithms.

Table 6. Compares the valid interval of parameter $\delta$ and parameter Eps between

ECNNI algorithm and DBSCAN algorithm

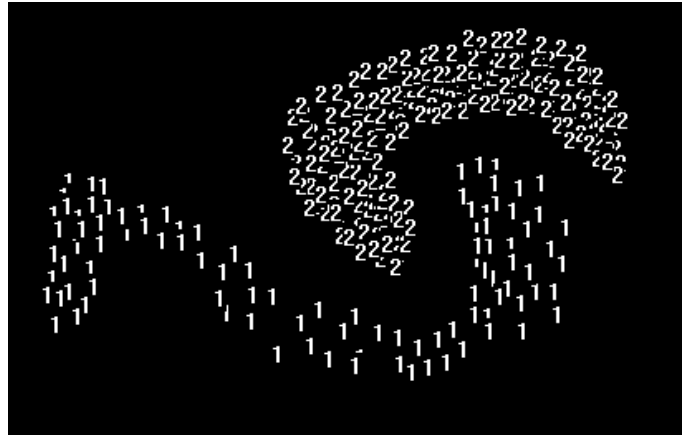

(a) Clusters of data1 identified by ECNNI and DBSCAN

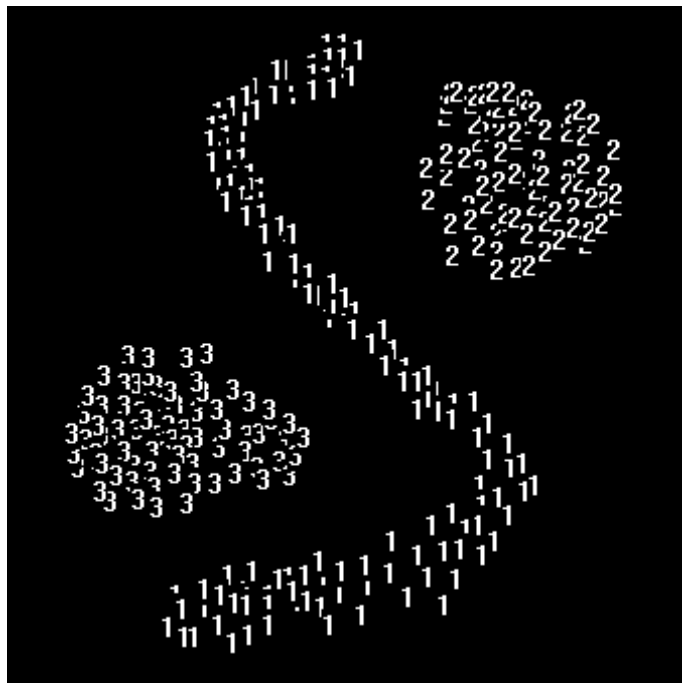

(b) Clusters of data2 identified by ECNNI and DBSCAN

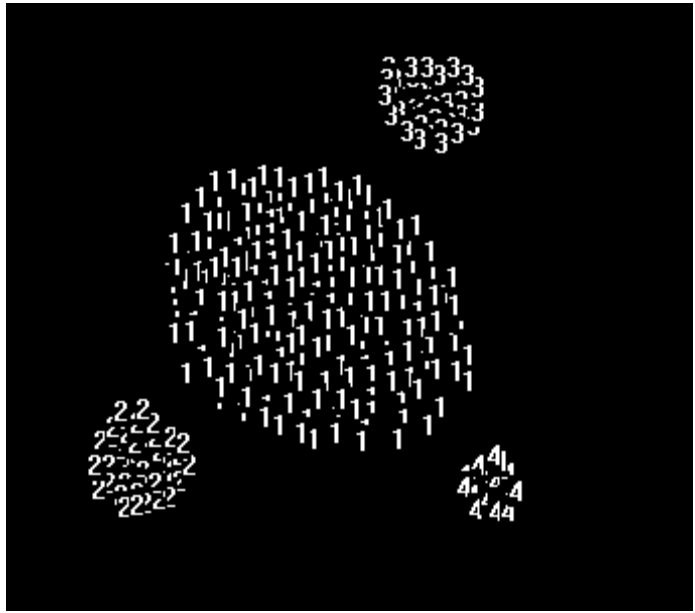

(c) Clusters of data3 identified by ECNNI and DBSCAN

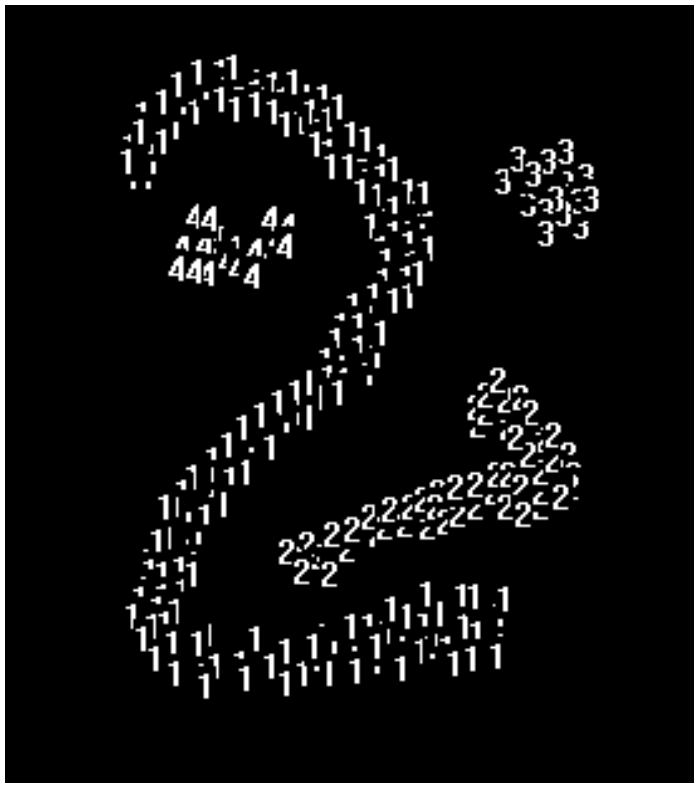

(d) Clusters of data4 identified by ECNNI and DBSCAN

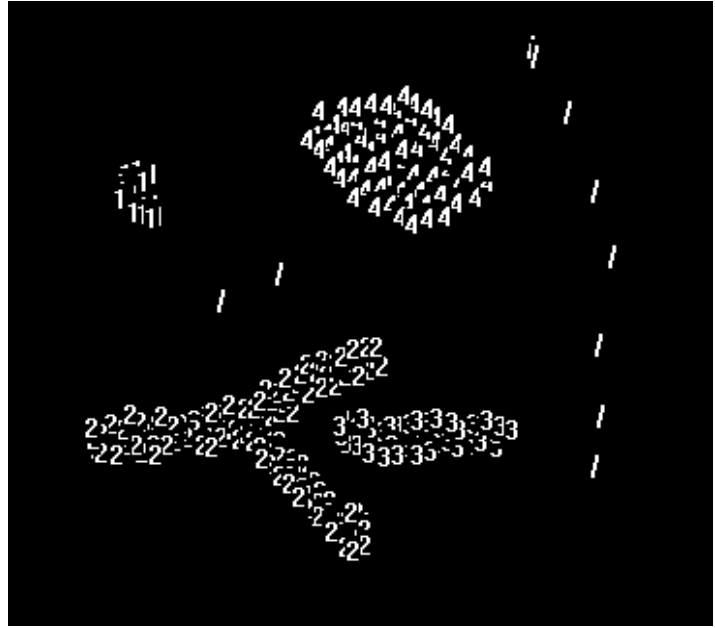

(e) Clusters of data5 identified by ECNNI and DBSCAN

Figure 3. the clustering results of Some Artificial Data Sets (data1-data5) using ECNNI and DBSCAN algorithms

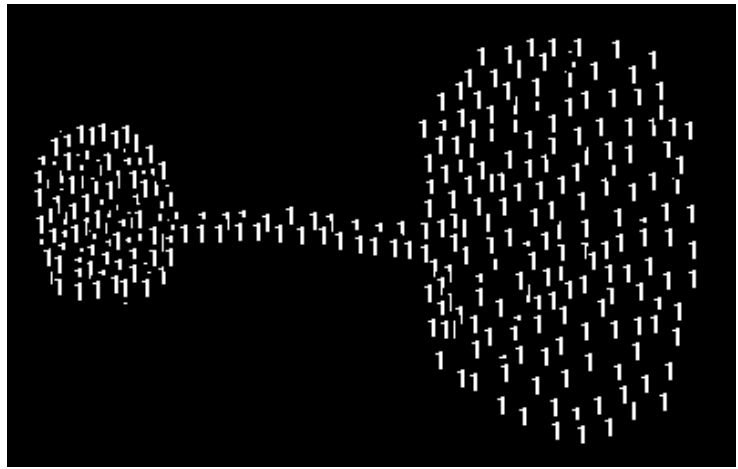

(a) Clusters of data6 identified by ECNNI (delta = 50) and DBSCAN (*MinPts* = 4, Eps = 50)

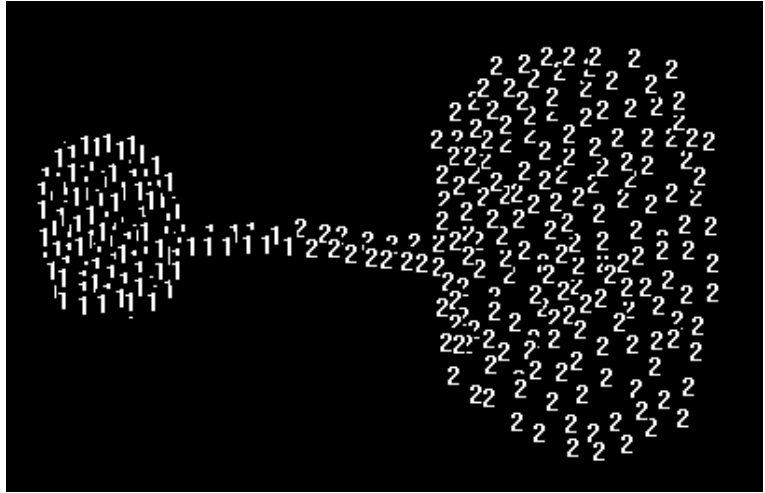

(b) Clusters of data6 identified by CNNI (delta = 50)

Figure 4. The obscure clustering results of an Artificial Data Sets (data6) using CNNI, ECNNI and DBSCAN algorithms

Table 6. Comparing the valid interval of parameter $\delta$ and parameter Eps between ECNNI algorithm and DBSCAN algorithm

|  | data1 | data2 | data3 | data4 | data5 |
|---|---|---|---|---|---|
| Valid interval of parameter Eps of DBSCAN (*MinPts* = 4) | [24, 31] | [17, 41] | [13, 28] | [13, 22] | [11, 18] |
| Valid interval of parameter $\delta$ of ECNNI | [24, 31] | [16, 41] | [13, 28] | [12, 22] | [11, 18] |

## 5.4 Compare the Clustering Purity with AP and KMeans algorithm Using Two UCI Data Sets

Table 7 lists the clustering results of CNNI algorithm and two comparative algorithms in time cost and clustering quality. From this experiment, we find that CNNI algorithm has variable clustering quality for variable value of parameter $\delta$, which means CNNI algorithm is sensitive to parameter $\delta$. CNNI algorithm can get better clustering results than AP or KMeans in some aspects.

Table 7. Comparison of Clustering results of CNNI algorithm and two comparative algorithms

| Data Sets | CNNI | | | KMeans | | | AP | | |
|---|---|---|---|---|---|---|---|---|---|
|  | *ST* | *NC* | *CP* | *ST* | *NC* (predefined) | *CP* | *ST* | *NC* | *CP* |
| Iris ($\delta$=0.8) | 0 | 4 | 92.67% | 0 | 3 | 88% | 5 | 11 | 94% |
| Wine ($\delta$=0.4) | 0 | 57 | 80.33% | 0 | 3 | 69.66% | 7 | 23 | 76.97% |

## 5.5 Validate the Compress Result of CNNI Algorithm by Clustering Pixel Points

**of Two Bmp Pictures in RGB Space**

Two bmp pictures are named as Picture1 and Picture2. Figure 5 gives the RGB space distribution of 200 * 200 pixel points in Picutre1. Figure 6. lists the original picture and five compressed pictures of Picture1. The five compressed pictures are drawn using the means of clusters obtained by clustering pixel points of Picutre1 in RGB space using CNNI algorithm. Figure 7 and Figure 8 are similar to Figure 5 and Figure 6 respectively.

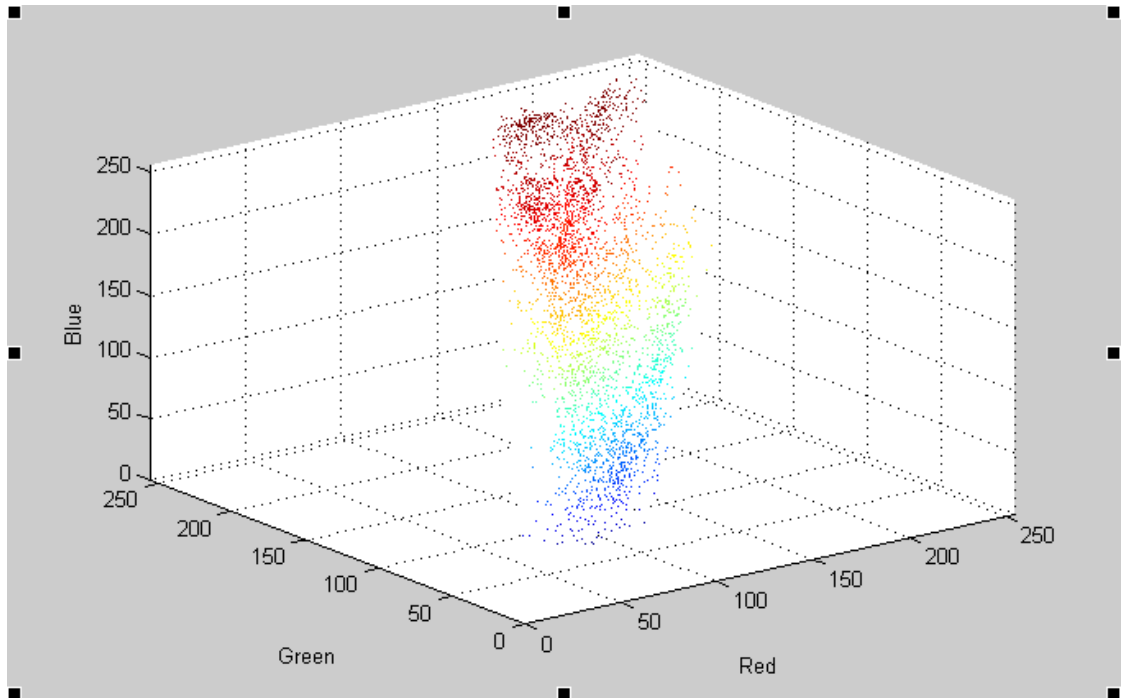

Figure 5. The RGB space distribution of 200 * 200 pixel points in Picutre1

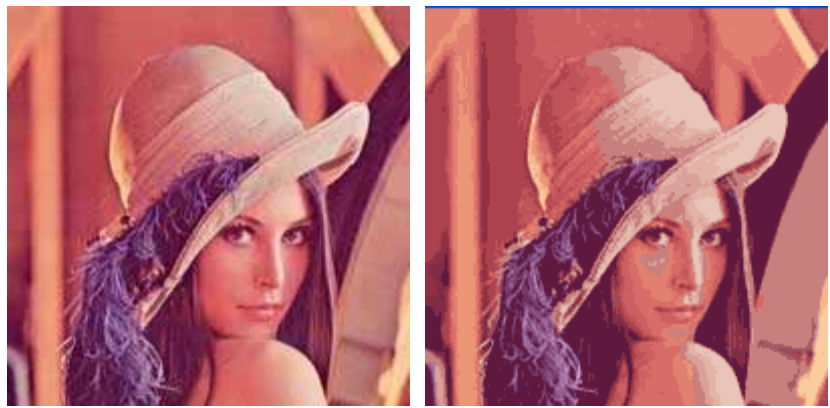

(a) The original Picture1          (b) $\delta$ = 12, CN = 32, ADM = 13.22

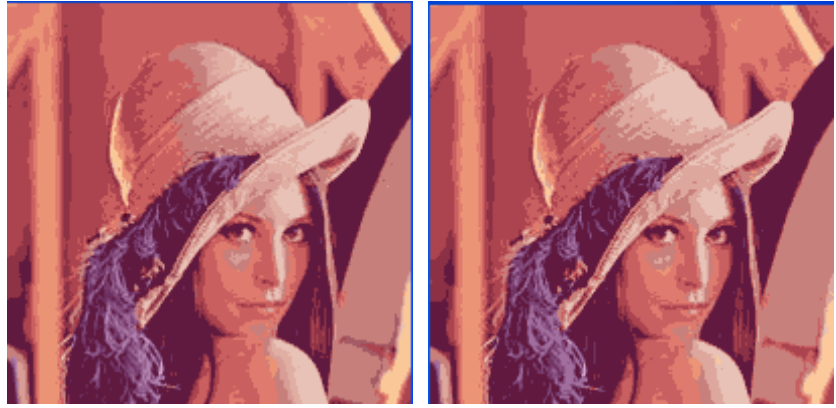

(c) δ = 13, CN = 23, ADM = 14.70  (d) δ = 16, CN = 16, ADM = 15.54

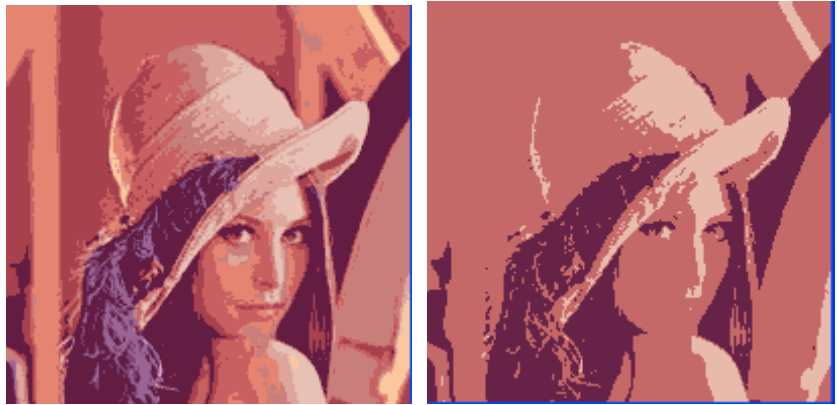

(e) δ = 20, CN = 11, ADM = 17.75  (f) δ = 32, CN = 3, ADM = 35.62

Figure 6. Five compressed pictures by clustering pixel points of Picutre1 in RGB space

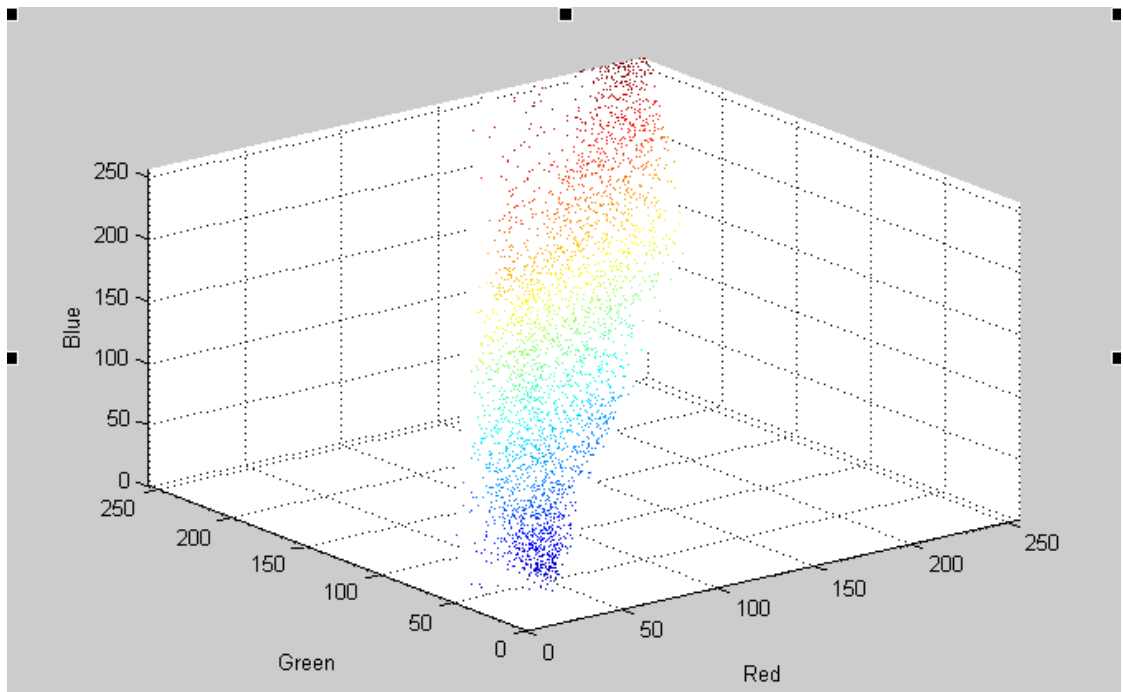

Figure 7. The RGB space distribution of 200 * 200 pixel points in Picture2

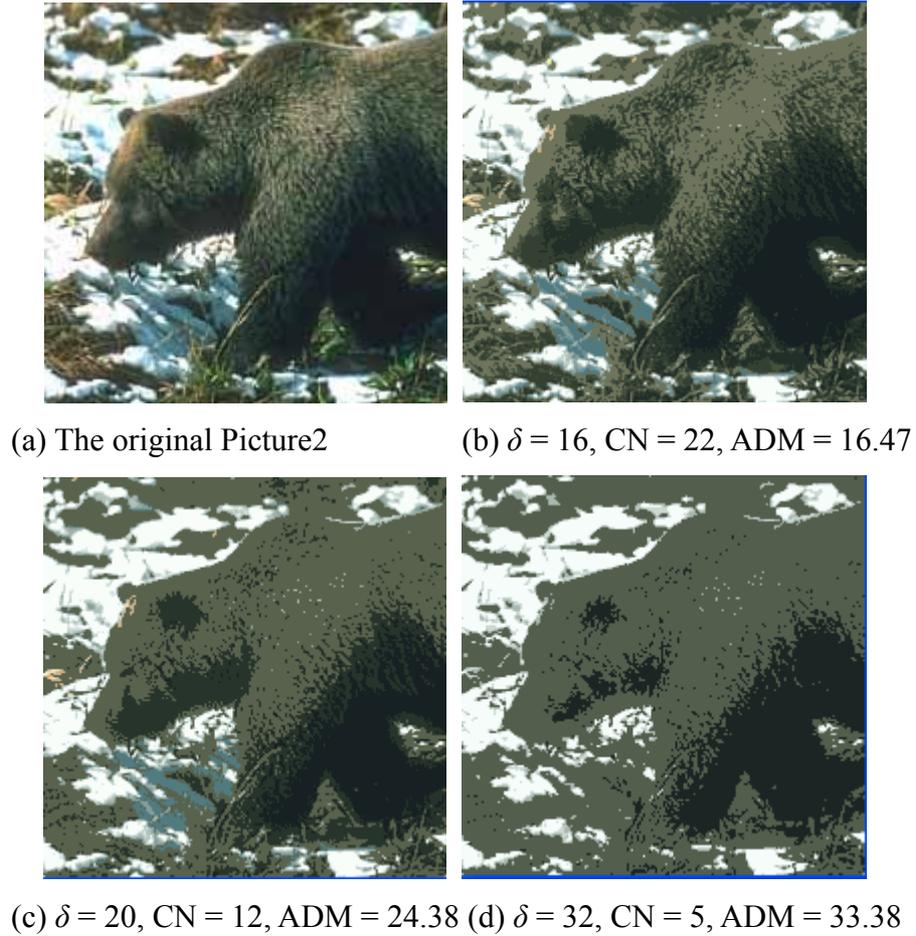

(a) The original Picture2      (b) $\delta = 16$, CN = 22, ADM = 16.47

(c) $\delta = 20$, CN = 12, ADM = 24.38   (d) $\delta = 32$, CN = 5, ADM = 33.38

Figure 8. Three compressed pictures by clustering pixel points of Picutre2 in RGB space

**5.6 Compare the Valid Interval of Parameter $\delta$ Between CNNI Algorithm and ECNNI Algorithm Using Some Artificial Data Sets (DS5 – DS8)**

Table 8 lists the valid interval of parameter $\delta$ of four artificial data sets between CNNI algorithm and ECNNI algorithm. Here, $[e_k , e_{k+1}]$ can be obtained from equation (5). From Table 8, we find that the valid interval of parameter $\delta$ in ECNNI algorithm is equal to $[\lceil e_k \rceil, \lfloor e_{k+1} \rfloor]$.

Figure 9 gives two comparison pictures of clustering results of DS7 ($n = 250$) corresponding to two different values of parameter $\delta$ by using ECNNI algorithm.

DS5: $[e_k , e_{k+1}]$ = [16.56, 42.02], ADM = 22.36;

DS6: $[e_k , e_{k+1}]$ = [13.41, 39.32], ADM = 16.39;

DS7: $[e_k , e_{k+1}]$ = [12.29, 49.10], ADM = 19.07;

DS8: $[e_k , e_{k+1}]$ = [13.41, 42.95], ADM = 16.18.

Table 8. Comparing the valid interval of parameter δ

between CNNI algorithm and ECNNI algorithm

| Valid interval of parameter δ | DS5 (*n* = 200) | | DS6 (*n* = 300) | | DS7 (*n* = 250) | | DS8 (*n* = 400) | |
|---|---|---|---|---|---|---|---|---|
| CNNI | δ | [27, 62] | δ | [21, 53] | δ | [24, 86] | δ | [24, 72] |
| ECNNI | δ | [17, 42] | δ | [14, 39] | δ | [13, 49] | δ | [14, 42] |

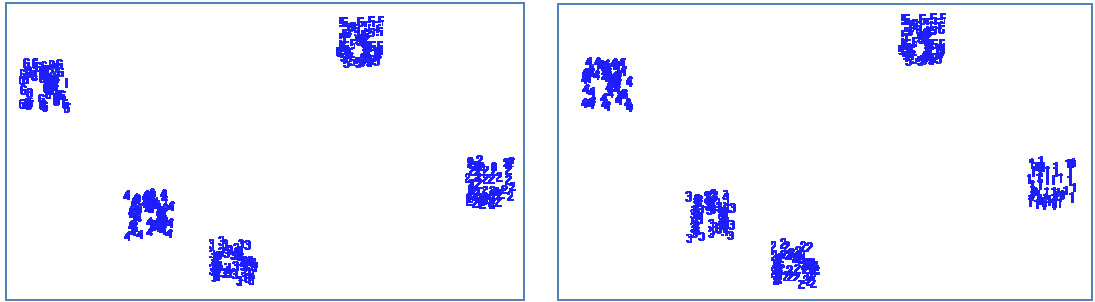

(a) δ = 12       (b) δ = 13

Figure 9. Comparison of clustering results of ECNNI Algorithm for two different values of parameter δ in DS7 (*n* = 250)

**5.7 Analysis and Conclusions of Experimental Results**

From Figure 1 to Figure 4, we find that CNNI, ICNNI, and ECNNI algorithm can find those obvious clusters of some artificial data sets.

From Table 4 and Table 5, ICNNI algorithm is validated in decreasing time cost. We find that constructing the δ near neighbor point sets is the main part of time cost.

From Figure 4, we find that ECNNI algorithm can get more obvious clusters than DBSCAN algorithm for some artificial data sets. From Table 6, we find that ECNNI algorithm has longer valid interval of parameter δ than parameter Eps of DBSCAN. DBSCAN algorithm needs two parameters, and ECNNI algorithm only needs one parameter.

From Figure 5 and Figure 7, we find that the clusters of Picture1 and Picture2 are not very obvious. This is the main reason that the experimental results of CNNI algorithm are sensitive to parameter δ.

From Table 8, we find that the valid interval of parameter δ in ECNNI algorithm is mainly less than the valid interval of parameter δ in CNNI algorithm.

CNNI algorithm is a new clustering algorithm with fast clustering speed. It can often get good clustering quality, although it is sensitive to parameter δ for some scatter data sets. ICNNI algorithm is an improved clustering algorithm with faster clustering speed than CNNI. ICNNI algorithm is sensitive to parameters δ and $r_i$ ($i = 1$, 2, …, $m$). Generally, $r_i$ ($i = 1, 2, …, m$) $\geq \delta$ is a better choice. ECNNI algorithm is an extended clustering algorithm with less valid interval of parameter $\delta$ than CNNI and ICNNI, which often results in faster speed in constructing $\delta$ near neighbor point sets.

## 6. Conclusions

This paper presents a new effective clustering algorithm CNNI, which can often get better clustering results or faster clustering speed for some data sets than some classical clustering algorithms.

CNNI algorithm is robust to outliers and allows to find clusters with different shapes. The number of clusters does not have to be fixed before clustering. Usually, parameter $\delta$ has some valid interval that can be determined by using an exploring method. In the process of constructing $\delta$ near neighbor point sets, if using an improved constructing method [1], or using the SS-tree [15] (or SR-tree [16]) index structure, the time complexity of this algorithm can even be decreased to O($n\log n$), which means it can be suitable to large data sets.

CNNI algorithm is sensitive to parameters $\delta$ when many noises and few obvious clusters exist, and it also cannot generate clusters with different levels of scatter because parameter $\delta$ is fixed before clustering.

The next work is to further improve this algorithm in time cost or clustering quality and overcome its two drawbacks.